\documentstyle[aps,psfig]{revtex}

\begin{document}

\nopagebreak
\title{\hspace{5.0in}Fermilab-Pub-98/275 \vspace{0.5in}
Formation of Patterns and Coherent Structures in Charged Particle Beams}
\author{~Stephan~I.~Tzenov}
\address{{\it Fermi National Accelerator Laboratory }\\
{\it P.~O.~Box 500, Batavia, IL 60510, USA}\\
{\it E.~mail: tzenov@fnal.gov}}
\maketitle

\newlength{\figwid} \newlength{\figlen} \medskip 

\begin{abstract}

In the present paper we study the long wavelength and slow time scale 
behavior of a coasting beam in a resonator adopting a broad-band impedance 
model. Based on the renormalization group approach we derive a set of 
coupled evolution equations for the beam envelope distribution function 
and the resonator voltage amplitude. The equation for the resonator voltage 
amplitude is further transformed into a generalized Ginzburg-Landau 
equation.

\end{abstract}

\section{Introduction.}

So far nonlinear wave phenomena have received scant attention in the study
of collective effects in charged particle beams. Considerable experimental
and simulation data however exists suggesting that these phenomena should be
included into the entire physical picture of beam propagation in
accelerators and storage rings.

A vast literature in the field of plasma physics is dedicated to the study
of nonlinear wave-particle processes due to space charge interparticle
forces. In high energy particle accelerators, where space charge forces are
negligibly small, of particular interest is the coherent state of the beam
under the influence of wakefields, or in the frequency domain, machine
impedance. This state is highly nonlinear and depends on the interaction of
nonlinear waves, involving some weak dissipative mechanisms that balance
beam fluctuations driven be the wakefields.

In a previous work \cite{tzenov}, \cite{colestock} we studied nonlinear
behavior of a coasting beam under the influence of a resonator impedance.
Starting from the gas-dynamic equations for longitudinal motion and using a
renormalization group (RG) approach \cite{goldenfeld}, \cite{kunihiro} we
found a set of coupled nonlinear equations for the beam density and
resonator voltage. However, as is well-known, the hydrodynamic approximation
is valid when the beam is close to a local equilibrium, which in a number of
practically important cases may well be far from reality.

The present paper, providing a complete kinetic description of the processes
involved, is aimed to overcome the above mentioned difficulties. In what
follows we study the longitudinal dynamics of a coasting beam in a resonator
adopting a broad-band impedance model. We are interested in describing slow
motion of beam patterns (droplets) neglecting fast oscillations of beam
density and voltage on the resonator at a frequency close to the resonant
frequency. We employ the RG method to derive amplitude equations governing
the dynamics of slow processes. In Section II we obtain the desired
equations for the longitudinal envelope distribution function and the
amplitude of the resonator voltage. In Section III we proceed to transform
the equation for the voltage amplitude into a generalized Ginzburg-Landau
equation by solving explicitly the Vlasov equation for the envelope
distribution function. Finally in Section IV we draw some conclusions
resulting from the work performed.

\section{The Amplitude Equations.}

The starting point for the subsequent analysis is the system of equations:

\[
\frac{\partial f}{\partial T}+v\frac{\partial f}{\partial \theta }+\lambda V%
\frac{\partial f}{\partial v}=0, 
\]

\begin{equation}
\frac{\partial ^2V}{\partial T^2}+2\gamma \frac{\partial V}{\partial T}%
+\omega ^2V=\frac{\partial I}{\partial T},  \label{kinetic}
\end{equation}

\[
I\left( \theta ;T\right) =\int dvvf\left( \theta ,v;T\right) , 
\]

\noindent for the longitudinal distribution function $f\left( \theta
,v;T\right) $ of an unbunched beam and the variation per turn of the voltage 
$V\left( \theta ;T\right) $ on a resonator. All dependent and independent
variables, as well as free parameters in equations (\ref{kinetic}) are
dimensionless and have been rescaled according to the relations:

\begin{equation}
T=\omega _st\quad ;\quad v=\frac 1{\omega _s}\frac{d\theta }{dt}=1+\frac{%
k_o\Delta E}{\omega _s}\quad ;\quad \omega =\frac{\omega _R}{\omega _s}\quad
;\quad \gamma =\frac \omega {2Q},
\end{equation}

\begin{equation}
\lambda =\frac{e^2R\gamma k_o\rho _o}\pi .
\end{equation}

\noindent Here $\omega _s$ is the angular revolution frequency of the
synchronous particle, $\Delta E$ is the energy error, $\omega _R$ is the
resonant frequency, $Q$ is the quality factor of the resonator, $R$ is the
resonator shunt impedance and $\rho _o$ is the uniform beam density
distribution at the thermodynamic limit. Furthermore

\begin{equation}
k_o=-\frac{\eta \omega _s}{\beta _s^2E_s}
\end{equation}

\noindent is the proportionality constant between the frequency deviation
and the energy deviation of a non synchronous particle with respect to the
synchronous one, while $\eta =\alpha _M-\gamma _s^{-2}$ ($\alpha _M$ -
momentum compaction factor) is the phase slip coefficient. The voltage
variation per turn $V,$ the beam current $I$ and the longitudinal
distribution function $f$ entering equations (\ref{kinetic}) have been
rescaled as well from their actual values $V_a$, $I_a$ and $f_a$ as 
follows:

\begin{equation}
V_a=2e\omega _s\rho _o\gamma RV\quad ;\quad I_a=e\omega _s\rho _oI\quad
;\quad f_a=\rho _of
\end{equation}

Let us introduce the Radon transform \cite{horvath}, \cite{klim} of the
distribution function $f\left( \theta ,v;T\right) $

\begin{equation}
f\left( \theta ,v;T\right) =\int d\xi F\left( \theta ,\xi ;T\right) \delta
\left[ v-U\left( \theta ,\xi ;T\right) \right] .  \label{radon}
\end{equation}

\noindent In the definition (\ref{radon}) $\xi $ can be viewed as a Lagrange
variable which is usually determined from the condition that the
distribution function $f\left( \theta ,v;T\right) $ be equal to a specified
distribution, say the equilibrium distribution for instance:

\[
f\left( \theta ,v;T\right) =f_0\left( \xi \right) \qquad \Rightarrow \qquad
v=U\left( \theta ,\xi ;T\right) . 
\]

\noindent Substitution of eq. (\ref{radon}), into the system (\ref{kinetic})
yields:

\[
\frac{\partial F}{\partial T}+\varepsilon \frac \partial {\partial \theta
}\left( FU\right) =0, 
\]

\begin{equation}
\frac{\partial U}{\partial T}+\varepsilon U\frac{\partial U}{\partial \theta 
}=\lambda V,  \label{hydrodyn}
\end{equation}

\[
\frac{\partial ^2V}{\partial T^2}+2\gamma \frac{\partial V}{\partial T}%
+\omega ^2V=\int d\xi \frac \partial {\partial T}\left( FU\right) , 
\]

\noindent where the fact that the azimuth $\theta $ is a slow variable (the
dependence of $F,$ $U$ and $V$ on $\theta $ is through a stretched variable $%
\zeta =\varepsilon \theta $) has been taken into account. Note that the system 
(\ref{hydrodyn}) resembles the set of gas-dynamic equations, governing the 
longitudinal motion of the beam. It bears however, additional information 
about the velocity distribution, embedded in the dependence on the Lagrange 
variable $\xi $, and takes into account its overall effect through the integral 
on the right hand side of the third equation.

We next examine the solution of the system of equations (\ref{hydrodyn}) order 
by order in the formal small parameter $\varepsilon $ by carrying out a naive
perturbation expansion. The zero order solution (stationary solution) is readily 
found to be:

\[
F_0=F_0\left( \xi \right) \quad ;\quad U_0=U_0\left( \xi \right) \quad
;\quad V_0\equiv 0 
\]

\noindent and in particular one can choose

\[
F_0\left( \xi \right) =f_0\left( \xi \right) \qquad ;\qquad U_0\left( \xi
\right) =1+\xi . 
\]

\noindent Combining the first order equations

\[
\frac{\partial F_1}{\partial T}=0\qquad ;\qquad \frac{\partial U_1}{\partial
T}=\lambda V_1, 
\]

\[
\frac{\partial ^2V_1}{\partial T^2}+2\gamma \frac{\partial V_1}{\partial T}%
+\omega ^2V_1=\int d\xi F_0\frac{\partial U_1}{\partial T}, 
\]

\noindent yields trivially a unique equation for $V_1$:

\[
\frac{\partial ^2V_1}{\partial T^2}+2\gamma \frac{\partial V_1}{\partial T}%
+\omega _o^2V_1=0\qquad ;\qquad \omega _o^2=\omega ^2-\lambda . 
\]

\noindent Solving the first order equations one easily obtains:

\[
V_1\left( \theta ;T\right) =E\left( \theta ;T_0\right) e^{i\omega _1\Delta
T}+c.c. 
\]

\begin{equation}
U_1\left( \theta ,\xi ;T\right) =u_o\left( \theta ,\xi ;T_0\right) +\lambda 
\frac{E\left( \theta ;T_0\right) }{i\omega _1}e^{i\omega _1\Delta T}+c.c.
\label{first}
\end{equation}

\[
F_1\left( \theta ,\xi ;T\right) =R_o\left( \theta ,\xi ;T_0\right) , 
\]

\noindent where

\begin{equation}
\omega _1=\omega _q+i\gamma \quad ;\quad \omega _q^2=\omega _o^2-\gamma
^2\quad ;\quad \Delta T=T-T_0,  \label{const}
\end{equation}

\noindent and the amplitudes $E\left( \theta ;T_0\right) ,$ $u_o\left(
\theta ,\xi ;T_0\right) ,$ $R_o\left( \theta ,\xi ;T_0\right) $ are yet
unknown functions of $\theta $, $\xi $ and the initial instant of time $T_0.$
Proceeding further with the expansion in the formal parameter $\varepsilon $
we write down the second order equations

\[
\frac{\partial F_2}{\partial T}+F_0\frac{\partial U_1}{\partial \theta }+U_0%
\frac{\partial R_o}{\partial \theta }=0, 
\]

\[
\frac{\partial U_2}{\partial T}+U_0\frac{\partial U_1}{\partial \theta }%
=\lambda V_2, 
\]

\[
\frac{\partial ^2V_2}{\partial T^2}+2\gamma \frac{\partial V_2}{\partial T}%
+\omega ^2V_2=\int d\xi \left( F_0\frac{\partial U_2}{\partial T}+\lambda
R_oV_1+U_0\frac{\partial F_2}{\partial T}\right) 
\]

\noindent and by elimination of $U_2$ and $F_2$ from the third equation we
obtain:

\[
\frac{\partial ^2V_2}{\partial T^2}+2\gamma \frac{\partial V_2}{\partial T}%
+\omega _o^2V_2=\lambda V_1\int d\xi R_o-2\int d\xi U_0F_0\frac{\partial
U_1}{\partial \theta }-\int d\xi U_0^2\frac{\partial R_o}{\partial \theta }%
. 
\]

\noindent Solving the above equation and subsequently the two other
equations for $U_2$ and $F_2$ we find the second order solution as follows:

\[
V_2\left( \theta ;T\right) =-\frac 1{\omega _o^2}\int d\xi \left( U_0^2%
\frac{\partial R_o}{\partial \theta }+2F_0U_0\frac{\partial u_o}{\partial
\theta }\right) + 
\]

\[
+\frac{\lambda \Delta T}{2i\omega _q}\left( E\int d\xi R_o+\frac{2i}{\omega
_1}\frac{\partial E}{\partial \theta }\int d\xi F_0U_0\right) e^{i\omega
_1\Delta T}+c.c. 
\]

\[
U_2\left( \theta ,\xi ;T\right) =-\Delta TU_0\frac{\partial u_o}{\partial
\theta }-\frac{\lambda \Delta T}{\omega _o^2}\int d\xi \left( U_0^2\frac{%
\partial R_o}{\partial \theta }+2F_0U_0\frac{\partial u_o}{\partial \theta }%
\right) + 
\]

\begin{equation}
+\frac{\lambda U_0}{\omega _1^2}\frac{\partial E}{\partial \theta }%
e^{i\omega _1\Delta T}-\frac{\lambda ^2\Delta T}{2\omega _1\omega _q}\left(
E\int d\xi R_o+\frac{2i}{\omega _1}\frac{\partial E}{\partial \theta }\int
d\xi F_0U_0\right) e^{i\omega _1\Delta T}+  \label{second}
\end{equation}

\[
+\frac{\lambda ^2}{2i\omega _q\omega _1^2}\left( E\int d\xi R_o+\frac{2i}{%
\omega _1}\frac{\partial E}{\partial \theta }\int d\xi F_0U_0\right)
e^{i\omega _1\Delta T}+c.c. 
\]

\[
F_2\left( \theta ,\xi ;T\right) =-\left( U_0\frac{\partial R_o}{\partial
\theta }+F_0\frac{\partial u_o}{\partial \theta }\right) \Delta T+\frac{%
\lambda F_0}{\omega _1^2}\frac{\partial E}{\partial \theta }e^{i\omega
_1\Delta T}+c.c. 
\]

\noindent In a way similar to the above we write the third order equations as

\[
\frac{\partial F_3}{\partial T}+F_0\frac{\partial U_2}{\partial \theta }%
+\frac \partial {\partial \theta }\left( F_1U_1\right) +U_0\frac{\partial F_2%
}{\partial \theta }=0, 
\]

\[
\frac{\partial U_3}{\partial T}+U_1\frac{\partial U_1}{\partial \theta }+U_0%
\frac{\partial U_2}{\partial \theta }=\lambda V_3, 
\]

\[
\frac{\partial ^2V_3}{\partial T^2}+2\gamma \frac{\partial V_3}{\partial T}%
+\omega ^2V_3=\int d\xi \left[ F_0\frac{\partial U_3}{\partial T}+R_o\frac{%
\partial U_2}{\partial T}+\frac{\partial \left( F_2U_1\right) }{\partial T}%
+U_0\frac{\partial F_3}{\partial T}\right] . 
\]

\noindent Solving the equation for $V_3$

\[
\frac{\partial ^2V_3}{\partial T^2}+2\gamma \frac{\partial V_3}{\partial T}%
+\omega _o^2V_3= 
\]

\[
=\int d\xi \left[ -2F_0U_1\frac{\partial U_1}{\partial \theta }-2F_0U_0%
\frac{\partial U_2}{\partial \theta }-2U_0\frac{\partial \left(
R_oU_1\right) }{\partial \theta }\right] + 
\]

\[
+\int d\xi \left( \lambda R_oV_2+\lambda V_1F_2-U_0^2\frac{\partial F_2}{%
\partial \theta }\right) 
\]

\noindent that can be obtained by combining the third order equations, and
subsequently solving the two other equations for $U_3$ and $F_3$ we obtain
the third order solution:

\[
V_3\left( \theta ;T\right) =\frac{\lambda \Delta T}{2i\omega _q}\left\{ 
\frac{2i}{\omega _1}\frac \partial {\partial \theta }\left[ E\int d\xi
\left( u_oF_0+U_0R_o\right) \right] -\right. 
\]

\[
-\frac 3{\omega _1^2}\left( \int d\xi F_0U_0^2\right) \frac{\partial ^2E}{%
\partial \theta ^2}+\frac 1{2i\omega _q}\left( 1-i\omega _q\Delta T\right)
\left[ \int d\xi \left( U_0\frac{\partial R_o}{\partial \theta }+F_0\frac{%
\partial u_o}{\partial \theta }\right) \right] E+ 
\]

\[
\left. +\frac \lambda {4\omega _q^2}\left( 1-i\omega _q\Delta T\right)
\left( \int d\xi R_o\right) \left[ E\int d\xi R_o+\frac{2i}{\omega _1}%
\left( \int d\xi F_0U_0\right) \frac{\partial E}{\partial \theta }\right]
\right\} e^{i\omega _1\Delta T}+ 
\]

\begin{equation}
+c.c.+\mathrm{oscillating\ terms}  \label{third}
\end{equation}

\[
U_3\left( \theta ,\xi ;T\right) =-u_o\frac{\partial u_o}{\partial \theta }%
\Delta T+\frac{\lambda ^2}{2\gamma \omega _o^2}\frac{\partial \left|
E\right| ^2}{\partial \theta }e^{-2\gamma \Delta T}+\mathrm{oscillating\
terms} 
\]

\[
F_3\left( \theta ,\xi ;T\right) =-\frac{\partial \left( R_ou_o\right) }{%
\partial \theta }\Delta T+\mathrm{oscillating\ terms} 
\]

\noindent Next we collect the secular terms that would contribute to the
amplitude equations when applying the RG procedure. Setting now $\varepsilon
=1$ we write down the part of the solution of the system (\ref{hydrodyn})
that has to be renormalized

\[
F_{RG}\left( \theta ,\xi ;T,T_0\right) =\widetilde{F}\left( \theta ,\xi
;T_0\right) -\Delta T\frac \partial {\partial \theta }\left[ \widetilde{F}%
\left( \theta ,\xi ;T_0\right) \widetilde{U}\left( \theta ,\xi ;T_0\right)
\right] , 
\]

\[
U_{RG}\left( \theta ,\xi ;T,T_0\right) =\widetilde{U}\left( \theta ,\xi
;T_0\right) -\Delta T\widetilde{U}\left( \theta ,\xi ;T_0\right) \frac
\partial {\partial \theta }\widetilde{U}\left( \theta ,\xi ;T_0\right) + 
\]

\[
+\frac{\lambda ^2}{2\gamma \omega _o^2}\frac{\partial \left| E\left( \theta
;T_0\right) \right| ^2}{\partial \theta }e^{-2\gamma \Delta T}, 
\]

\begin{equation}
V_{RG}\left( \theta ;T,T_0\right) =\left\{ E+\frac{\lambda \Delta T}{%
2i\omega _q}\left[ E\int d\xi R_o+\frac{2i}{\omega _1}\frac \partial
{\partial \theta }\left( E\int d\xi \widetilde{F}\widetilde{U}\right)
-\right. \right.  \label{rgperturb}
\end{equation}

\[
-\frac 3{\omega _1^2}\left( \int d\xi F_0U_0^2\right) \frac{\partial ^2E}{%
\partial \theta ^2}-\frac{iE}{2\omega _q}\frac \partial {\partial \theta
}\left( \int d\xi \widetilde{F}\widetilde{U}\right) + 
\]

\[
\left. \left. +\frac \lambda {4\omega _q^2}\left( \int d\xi R_o\right)
\left( E\int d\xi R_o+\frac{2i}{\omega _1}\frac{\partial E}{\partial \theta 
}\int d\xi F_0U_0\right) \right] \right\} e^{i\omega _1\Delta T}+c.c. 
\]

\noindent where

\[
\widetilde{F}\left( \theta ,\xi ;T_0\right) =F_0\left( \xi \right)
+R_o\left( \theta ,\xi ;T_0\right) \quad ;\quad \widetilde{U}\left( \theta
,\xi ;T_0\right) =U_0\left( \xi \right) +u_o\left( \theta ,\xi ;T_0\right) . 
\]

\noindent Following Kunihiro \cite{kunihiro} we represent the solution (\ref
{rgperturb}) as a family of trajectories or curves $\left\{ \Re
_{T_0}\right\} =\left[ F_{RG}\left( T_0\right) ,U_{RG}\left( T_0\right)
,V_{RG}\left( T_0\right) \right] $, being parameterized with $T_0$. The RG
equations are defined as the envelope equations for the one-parameter family 
$\left\{ \Re _{T_0}\right\} $:

\begin{equation}
\left. \left( \frac{\partial F_{RG}}{\partial T_0},\frac{\partial U_{RG}}{%
\partial T_0},\frac{\partial V_{RG}}{\partial T_0}\right) \right| _{T_0=T}=0.
\label{envequation}
\end{equation}

\noindent From the above definition (\ref{envequation}) it is straightforward
to obtain the desired RG equations:

\[
\frac{\partial \widetilde{F}}{\partial T}+\frac \partial {\partial \theta
}\left( \widetilde{F}\widetilde{U}\right) =0, 
\]

\[
\frac{\partial \widetilde{U}}{\partial T}+\widetilde{U}\frac{\partial 
\widetilde{U}}{\partial \theta }=-\frac{\lambda ^2}{\omega _o^2}\frac{%
\partial \left| \widetilde{E}\right| ^2}{\partial \theta }, 
\]

\begin{equation}
\frac{2i\omega _q}\lambda \left( \frac \partial {\partial T}+\frac \partial
{\partial \theta }+ \gamma \right) \widetilde{E}=\widetilde{E}\int d\xi
\left( \widetilde{F}-F_0\right) +\frac{2i}{\omega _1}\frac \partial
{\partial \theta }\left( \widetilde{E}\int d\xi \widetilde{F}\widetilde{U}%
\right) -  \label{rgequations}
\end{equation}

\[
-\frac 3{\omega _1^2}\left( \int d\xi F_0U_0^2\right) \frac{\partial ^2%
\widetilde{E}}{\partial \theta ^2}-\frac i{2\omega _q}\widetilde{E}\frac
\partial {\partial \theta }\left( \int d\xi \widetilde{F}\widetilde{U}%
\right) + 
\]

\[
+\frac \lambda {4\omega _q^2}\left[ \int d\xi \left( \widetilde{F}%
-F_0\right) \right] \left[ \widetilde{E}\int d\xi \left( \widetilde{F}%
-F_0\right) +\frac{2i}{\omega _1}\frac \partial {\partial \theta }\left( 
\widetilde{E}\int d\xi \widetilde{F}\widetilde{U}\right) \right] , 
\]

\noindent where

\[
\widetilde{E}\left( \theta ;T\right) =E\left( \theta ;T\right) 
e^{- \gamma T}. 
\]

The final step consists in defining the envelope distribution function $%
G\left( \theta ,v;T\right) $ by the Radon transform

\begin{equation}
G\left( \theta ,v;T\right) =\int d\xi \widetilde{F}\left( \theta ,\xi
;T\right) \delta \left[ v-\widetilde{U}\left( \theta ,\xi ;T\right) \right] .
\label{radonenvel}
\end{equation}

\noindent By virtue of (\ref{radonenvel}) the system of RG equations (\ref
{rgequations}) is equivalent to the following system of equations for the
envelope distribution function $G\left( \theta ,v;T\right) $ and the
resonator voltage amplitude $\widetilde{E}\left( \theta ;T\right) $:

\begin{equation}
\frac{\partial G}{\partial T}+v\frac{\partial G}{\partial \theta }-\frac{%
\lambda ^2}{\omega _o^2}\frac{\partial \left| \widetilde{E}\right| ^2}{%
\partial \theta }\frac{\partial G}{\partial v}=0,  \label{vlasov}
\end{equation}

\[
\frac{2i\omega _q}\lambda \left( \frac \partial {\partial T}+\frac \partial
{\partial \theta }+ \gamma \right) \widetilde{E}=\widetilde{E}\int dv\left(
G-f_0\right) +\frac{2i}{\omega _1}\frac \partial {\partial \theta }\left( 
\widetilde{E}\int dvvG\right) - 
\]

\[
-\frac 3{\omega _1^2}\left[ \int dvv^2f_0\left( v\right) \right] \frac{%
\partial ^2\widetilde{E}}{\partial \theta ^2}-\frac i{2\omega _q}\widetilde{E%
}\frac \partial {\partial \theta }\left( \int dvvG\right) + 
\]

\begin{equation}
+\frac \lambda {4\omega _q^2}\left[ \int dv\left( G-f_0\right) \right]
\left[ \widetilde{E}\int dv\left( G-f_0\right) +\frac{2i}{\omega _1}\frac
\partial {\partial \theta }\left( \widetilde{E}\int dvvG\right) \right] .
\label{resonator}
\end{equation}

The system of equations (\ref{vlasov}) and (\ref{resonator}) provides a
complete description of nonlinear particle-wave interaction. It governs slow
processes of beam pattern dynamics through the evolution of the amplitude
functions. In (\ref{vlasov}) one can immediately recognize the Vlasov
equation for the envelope distribution function $G\left( \theta ,v;T\right) $
with the ponderomotive force, due to fast oscillations at frequency close to
the resonant frequency. It may be worth noting that the system (\ref{vlasov}%
) and (\ref{resonator}) intrinsically contains the nonlinear Landau damping
mechanism, a fact that will become apparent from the treatment in the next
Section.

\section{Derivation of the Generalized Ginzburg-Landau Equation.}

In order to solve equation (\ref{vlasov}) we perform a Fourier transform and
obtain:

\[
\left( \Omega -kv\right) G\left( \chi \right) = 
\]

\begin{equation}
-\frac{\lambda ^2}{\left( 2\pi \right) ^4\omega _o^2}\int d\chi _1d\chi
_2d\chi _3\delta \left( \chi -\chi _1-\chi _2-\chi _3\right) \left(
k-k_1\right) \frac{\partial G\left( \chi _1\right) }{\partial v}\widetilde{E}%
\left( \chi _2\right) \widetilde{E}^{*}\left( \chi _3\right) ,
\label{fourier}
\end{equation}

\noindent where

\[
\chi =\left( k,\Omega \right) \qquad ;\qquad \delta \left( \chi \right)
=\delta \left( k\right) \delta \left( \Omega \right) 
\]

\noindent and the Fourier transform of a generic function $g\left( \theta
;T\right) $ is defined as

\[
g\left( \theta ;T\right) =\frac 1{\left( 2\pi \right) ^2}\int d\Omega
dkg\left( k;\Omega \right) e^{i\left( k\theta -\Omega T\right) }, 
\]

\[
g\left( k;\Omega \right) =\int d\theta dTg\left( \theta ;T\right)
e^{-i\left( k\theta -\Omega T\right) }, 
\]

\[
g^{*}\left( k;\Omega \right) =\left[ g\left( -k;-\Omega \right) \right]
^{*}. 
\]

\noindent Solving equation (\ref{fourier}) perturbatively we represent its
solution in the form:

\begin{equation}
G\left( \chi \right) =\left( 2\pi \right) ^2f_0\left( v\right) \delta \left(
\chi \right) +\widetilde{G}\left( \chi \right) \quad ;\quad \widetilde{G}%
\left( \chi \right) =\sum\limits_{n=1}^\infty G_n\left( \chi \right) ,
\label{gensol}
\end{equation}

\noindent where

\begin{equation}
G_1\left( \chi \right) =-\frac{\lambda ^2}{\left( 2\pi \right) ^2\omega _o^2}%
\frac k{\Omega -kv}\frac{\partial f_0}{\partial v}\int d\chi _1d\chi
_2\delta \left( \chi -\chi _1-\chi _2\right) \widetilde{E}\left( \chi
_1\right) \widetilde{E}^{*}\left( \chi _2\right) ,  \label{sol1}
\end{equation}

\[
G_n\left( \chi \right) =-\frac{\lambda ^2}{\left( 2\pi \right) ^4\omega _o^2}%
\frac 1{\Omega -kv}* 
\]

\begin{equation}
\ast \int d\chi _1d\chi _2d\chi _3\delta \left( \chi -\chi _1-\chi _2-\chi
_3\right) \left( k-k_1\right) \frac{\partial G_{n-1}\left( \chi _1\right) }{%
\partial v}\widetilde{E}\left( \chi _2\right) \widetilde{E}^{*}\left( \chi
_3\right) .  \label{soln}
\end{equation}

\noindent The Fourier transform of equation (\ref{resonator}) yields the
linear dispersion relation

\[
\Omega =-i\gamma +k-\frac{\lambda \left\langle v\right\rangle _0}{\omega
_q\omega _1}k+\frac{3\lambda \left\langle v^2\right\rangle _0}{2\omega
_q\omega _1}k^2= 
\]

\begin{equation}
=k-\frac{\lambda \left\langle v\right\rangle _0}{\omega _o^2}k+\frac{%
3\lambda \left\langle v^2\right\rangle _0}{2\omega _q\omega _o^2}k^2-i\gamma
\left( 1-\frac{\lambda \left\langle v\right\rangle _0}{\omega _q\omega _o^2}%
k+\frac{3\lambda \left\langle v^2\right\rangle _0}{\omega _o^4}k^2\right) .
\label{dispersion}
\end{equation}

\noindent The integrals over $v$ of the envelope distribution function $%
G\left( \theta ,v;T\right) $

\[
I_0\left( \theta ;T\right) =\int dvG\left( \theta ,v;T\right) =\frac
1{\left( 2\pi \right) ^2}\int dvd\Omega dkG\left( k,v;\Omega \right)
e^{i\left( k\theta -\Omega T\right) }, 
\]

\[
I_1\left( \theta ;T\right) =\int dvvG\left( \theta ,v;T\right) =\frac
1{\left( 2\pi \right) ^2}\int dvd\Omega dkvG\left( k,v;\Omega \right)
e^{i\left( k\theta -\Omega T\right) } 
\]

\noindent entering equation (\ref{resonator}) can be computed in a
straightforward manner. Substituting the solution (\ref{gensol})-(\ref{soln}%
) into the above equations with the linear dispersion relation (\ref
{dispersion}) in hand, up to second order in $G\left( k,v;\Omega \right) $,
we find

\begin{equation}
I_0\left( \theta ;T\right) =1-W\left[ \left| \widetilde{E}\left( \theta
;T\right) \right| \right] +...,  \label{i0}
\end{equation}

\begin{equation}
I_1\left( \theta ;T\right) =1-\left( 1-\frac \lambda {\omega _o^2}\right)
W\left[ \left| \widetilde{E}\left( \theta ;T\right) \right| \right] +...,
\label{i1}
\end{equation}

\noindent where the function $W$ is defined as

\begin{equation}
W\left( z\right) =\frac{\lambda ^2}{\omega _o^2\sigma _v^2}\left( 1+i\gamma
_L\right) \left( 1-\frac{\lambda ^2}{2\omega _o^2\sigma _v^2}z^2\right) z^2,
\end{equation}

\noindent and

\begin{equation}
\gamma _L=\frac \lambda {\omega _o^2\sigma _v}\sqrt{\frac \pi 2}\exp \left( -%
\frac{\lambda ^2}{2\omega _o^4\sigma _v^2}\right)
\end{equation}

\noindent is the Landau damping factor. In the above calculations the
equilibrium distribution function has been taken to be the Gaussian one

\[
f_0\left( v\right) =\frac 1{\sigma _v\sqrt{2\pi }}\exp \left[ -\frac{\left(
v-1\right) ^2}{2\sigma _v^2}\right] , 
\]

\noindent where

\[
\sigma _v=\frac{\left| k_o\right| \sigma _E}{\omega _s} 
\]

\noindent and $\sigma _E$ is the r.m.s. of the energy error, proportional to
the longitudinal beam temperature. By substitution of the expressions (\ref
{i0}) and (\ref{i1}) into equation (\ref{resonator}) we arrive at the
generalized Ginzburg-Landau equation:

\[
\frac{2i\omega _q}\lambda \left( \frac \partial {\partial T}+\frac \partial
{\partial \theta }+ \gamma \right) \widetilde{E}=-\frac{3\left( 1+\sigma
_v^2\right) }{\omega _1^2}\frac{\partial ^2\widetilde{E}}{\partial \theta ^2}%
+\frac{2i}{\omega _1}\frac{\partial \widetilde{E}}{\partial \theta }-\left(
1-\frac \lambda {4\omega _q^2}W\right) W\widetilde{E}- 
\]

\[
-\frac{2i}{\omega _1}\frac \lambda {4\omega _q^2}W\frac{\partial \widetilde{E%
}}{\partial \theta }+\frac i{2\omega _q}\left( 1-\frac \lambda {\omega
_o^2}\right) \widetilde{E}\frac{\partial W}{\partial \theta }-\frac{2i}{%
\omega _1}\left( 1-\frac \lambda {\omega _o^2}\right) \left( 1-\frac \lambda
{4\omega _q^2}W\right) \frac \partial {\partial \theta }\left( W\widetilde{E}%
\right) . 
\]

\noindent It can be further cast to a simpler form by introducing the rescaled
independent and dependent variables according to

\[
\tau =\frac{\lambda T}{2\omega _q}\quad ;\quad x=\frac{\omega _o}{\sqrt{%
3\left( 1+\sigma _v^2\right) }}\left( \theta -T+\frac{\lambda T}{\omega
_o\omega _q}\right) \quad ;\quad \Psi =\frac{\left| \lambda \right| 
\widetilde{E}}{\omega _o\sigma _v}. 
\]

\noindent Then the generalized Ginzburg-Landau equation for the amplitude of
the resonator voltage takes its final form:

\[
i\frac{\partial \Psi }{\partial \tau } + ib \gamma \Psi =-\left( 1-\frac{%
2i\gamma }{\omega _o}\right) \frac{\partial ^2\Psi }{\partial x^2}+a\gamma 
\frac{\partial \Psi }{\partial x}-\left( 1-b_1W\right) W\Psi - 
\]

\begin{equation}
-ab_1\left( \gamma +i\omega _o\right) W\frac{\partial \Psi }{\partial x}%
-a_1\left( \gamma +i\omega _o\right) \left( 1-b_1W\right) \frac \partial
{\partial x}\left( W\Psi \right) +\frac{ia_1b_1b\omega _o^2}2\Psi \frac{%
\partial W}{\partial x},  \label{gglequat}
\end{equation}

\noindent where

\[
a=\frac 2{\omega _o\sqrt{3\left( 1+\sigma _v^2\right) }}\qquad ;\qquad b=%
\frac{2\omega _q}\lambda , 
\]

\[
b_1=\frac 1{\lambda b^2}\qquad ;\qquad a_1=a\left( 1-4b_1\right) , 
\]

\noindent and the function $W\left( \left| \Psi \right| \right) $ is given
now by the simple expression:

\begin{equation}
W\left( \left| \Psi \right| \right) =\left( 1+i\gamma _L\right) \left(
1-\frac 12\left| \Psi \right| ^2\right) \left| \Psi \right| ^2.
\label{nlpotent}
\end{equation}

The generalized Ginzburg-Landau equation (\ref{gglequat}) is known \cite
{cross} to provide the basic framework for the study of many properties of
non equilibrium systems, such as existence and interaction of coherent
structures, generic onset of travelling wave disturbance in continuous
media, appearance of chaos. Recent experimental and numerical evidence 
(see e.g. \cite{tzenov}, \cite{colestock} and the references therein) shows 
that similar behavior is consistent with the propagation of charged particle 
beams, and the generalized Ginzburg-Landau equation we have derived could 
represent the appropriate analytical model to study the above mentioned 
phenomena.

\section{Concluding Remarks.}

As a result of the investigation performed we have shown that a coasting
beam under the influence of a resonator impedance exhibits spatial-temporal
patterns modulated by an envelope (amplitude) function, which varies slowly
compared to the fast time and short wavelength scales of the pattern itself.
Extracting long wavelength and slow time scale behavior of the system we
have derived a set of coupled nonlinear evolution equations for the beam
envelope distribution function and voltage amplitude. We have further shown
that the amplitude of the nonlinear wave satisfies a generalized
Ginzburg-Landau equation.

It is worthwhile to mention that the analytical framework presented here
bears rather general features. It provides complete kinetic description of
slow, fully nonlinear particle-wave interaction process, and allows higher
order corrections to the generalized Ginzburg-Landau equation to be taken
into account.

\section{Acknowledgements.}

The author is indebted to Pat Colestock for many helpful discussions
concerning the subject of the present paper, and to David Finley and Steve
Holmes for their enlightened support of this work.

\end{document}